\documentclass[a4paper,11pt]{article}
\usepackage{pos}

\usepackage{graphicx}
\graphicspath{{figures/}}

\usepackage{float}
\usepackage{subfig}
\usepackage{multicol}
\usepackage{array}

\usepackage{hyperref}

\title{Double cascade reconstruction in the Baikal-GVD neutrino telescope}
 \ShortTitle{Double cascade reconstruction in Baikal-GVD}

\author[a]{V.M.~Aynutdinov}
\author[b]{V.A.~Allakhverdyan}
\author[a]{A.D.~Avrorin}
\author[a]{A.V.~Avrorin}
\author[c,d]{Z.~Barda\v{c}ov\'{a}}
\author[b]{I.A.~Belolaptikov}
\author[a]{E.A.~Bondarev}
\author[b]{I.V.~Borina}
\author[e]{N.M.~Budnev}
\author[l]{V.A.~Chadymov}
\author[f]{A.S.~Chepurnov}
\author[b,g]{V.Y.~Dik}
\author[a]{G.V.~Domogatsky}
\author[a]{A.A.~Doroshenko}
\author[c]{R.~Dvornick\'{y}}
\author[e]{A.N.~Dyachok}
\author[a]{Zh.-A.M.~Dzhilkibaev}
\author*[c,d]{E.~Eckerov\'{a}}
\author[b]{T.V.~Elzhov}
\author[d]{L.~Fajt}
\author[l]{V.N. Fomin}
\author[e]{A.R.~Gafarov}
\author[a]{K.V.~Golubkov}
\author[b]{N.S.~Gorshkov}
\author[e]{T.I.~Gress}
\author[h]{K.G.~Kebkal}
\author[a]{I.V.~Kharuk}
\author[b]{E.V.~Khramov}
\author[b]{M.M.~Kolbin}
\author[i]{S.O.~Koligaev}
\author[b]{K.V.~Konischev}
\author[b]{A.V.~Korobchenko}
\author[a]{A.P.~Koshechkin}
\author[f]{V.A.~Kozhin}
\author[b]{M.V.~Kruglov}
\author[j]{V.F.~Kulepov}
\author[e]{Y.E.~Lemeshev}
\author[a,\dagger]{M.B.~Milenin}
\author[e]{R.R.~Mirgazov}
\author[b]{D.V.~Naumov}
\author[f]{A.S.~Nikolaev}
\author[a]{D.P.~Petukhov}
\author[b]{E.N.~Pliskovsky}
\author[k]{M.I.~Rozanov}
\author[e]{E.V.~Ryabov}
\author[a]{G.B.~Safronov}
\author[b,g]{D.~Seitova}
\author[b]{B.A.~Shaybonov}
\author[a]{M.D.~Shelepov}
\author[a]{S.D.~Shilkin}
\author[f]{E.V.~Shirokov}
\author[c,d]{F.~\v{S}imkovic}
\author[b]{A.E.~Sirenko}
\author[f]{A.V.~Skurikhin}
\author[b]{A.G.~Solovjev}
\author[b]{M.N.~Sorokovikov}
\author[d]{I.~\v{S}tekl}
\author[a]{A.P.~Stromakov}
\author[a]{O.V.~Suvorova}
\author[e]{V.A.~Tabolenko}
\author[b]{B.B.~Ulzutuev}
\author[b]{Y.V.~Yablokova}
\author[a]{D.N.~Zaborov}
\author[b]{S.I.~Zavyalov}
\author[b]{D.Y.~Zvezdov}

\affiliation[a]{Institute for Nuclear Research, Russian Academy of Sciences, Moscow, 117312, Russia}
\affiliation[b]{Joint Institute for Nuclear Research, Dubna, 141980, Russia}
\affiliation[c]{Comenius University, 81499 Bratislava, Slovakia}
\affiliation[d]{Czech Technical University in Prague, Institute of Experimental and Applied Physics, 11000 Prague, Czech Republic}
\affiliation[e]{Irkutsk State University, Irkutsk, 664003, Russia}
\affiliation[f]{Skobeltsyn Institute of Nuclear Physics, Moscow State University, Moscow, 119991, Russia}
\affiliation[g]{Institute of Nuclear Physics of the Ministry of Energy of the Republic of Kazakhstan, Almaty, 050032, Kazakhstan}
\affiliation[h]{LATENA, St. Petersburg, 199106, Russia}
\affiliation[i]{INFRAD, Dubna, 141981, Russia}
\affiliation[j]{Nizhny Novgorod State Technical University, Nizhny Novgorod, 603950, Russia}
\affiliation[k]{St.~Petersburg State Marine Technical University, St.~Petersburg, 190008, Russia}
\affiliation[l]{Moscow, free researcher}

\note[\textdagger]{Deceased.}

\emailAdd{eliska.eckerova@fmph.uniba.sk}

\abstract{Baikal Gigaton Volume Detector is a cubic kilometer scale neutrino telescope under construction in Lake Baikal. As of July 2023, Baikal-GVD consists of 96 fully deployed strings resulting in 3456 optical modules installed. The observation of neutrinos is based on detection of Cherenkov radiation emitted by the products of neutrino interactions. In this contribution, description of the double cascade reconstruction technique as well as evaluation of precision of this algorithm is given.}

\ConferenceLogo{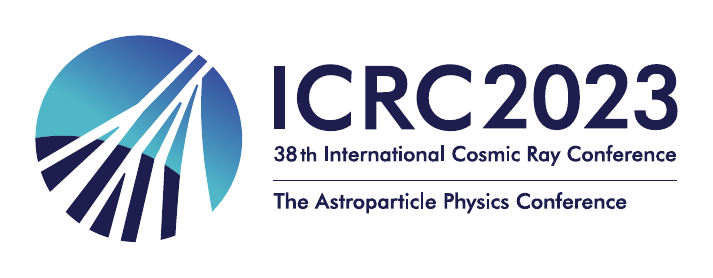}

\FullConference{%
The 38th International Cosmic Ray Conference (ICRC2023)\\
  26 July -- 3 August, 2023\\
  Nagoya, Japan}

\begin{document}
\maketitle

\section{Introduction}
The Baikal Gigaton Volume Detector (Baikal-GVD)~\cite{baikal_gvd} is a water Cherenkov neutrino telescope under construction in the southern part of Lake Baikal -- the deepest freshwater lake in the world. It is located approximately 3-4~km from the shore, where the lake bottom is relatively constant at $\sim$1366~m. Baikal-GVD is a three-dimensional grid of photomultiplier tubes aiming to detect Cherenkov radiation emitted by the charged particles originating in interactions of high-energy astrophysical neutrinos.

\begin{figure}[h!]
	\centering
	\begin{multicols}{2}
		\centering
		\subfloat[]{
			\includegraphics[width=1.1\linewidth]{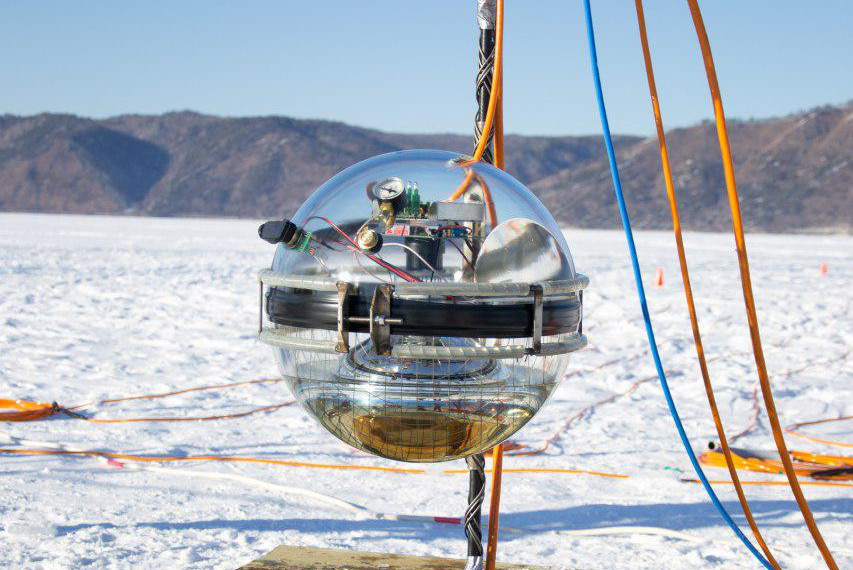}
			\label{obr:OM}
		}  
		\subfloat[]{
			\includegraphics[width=\linewidth]{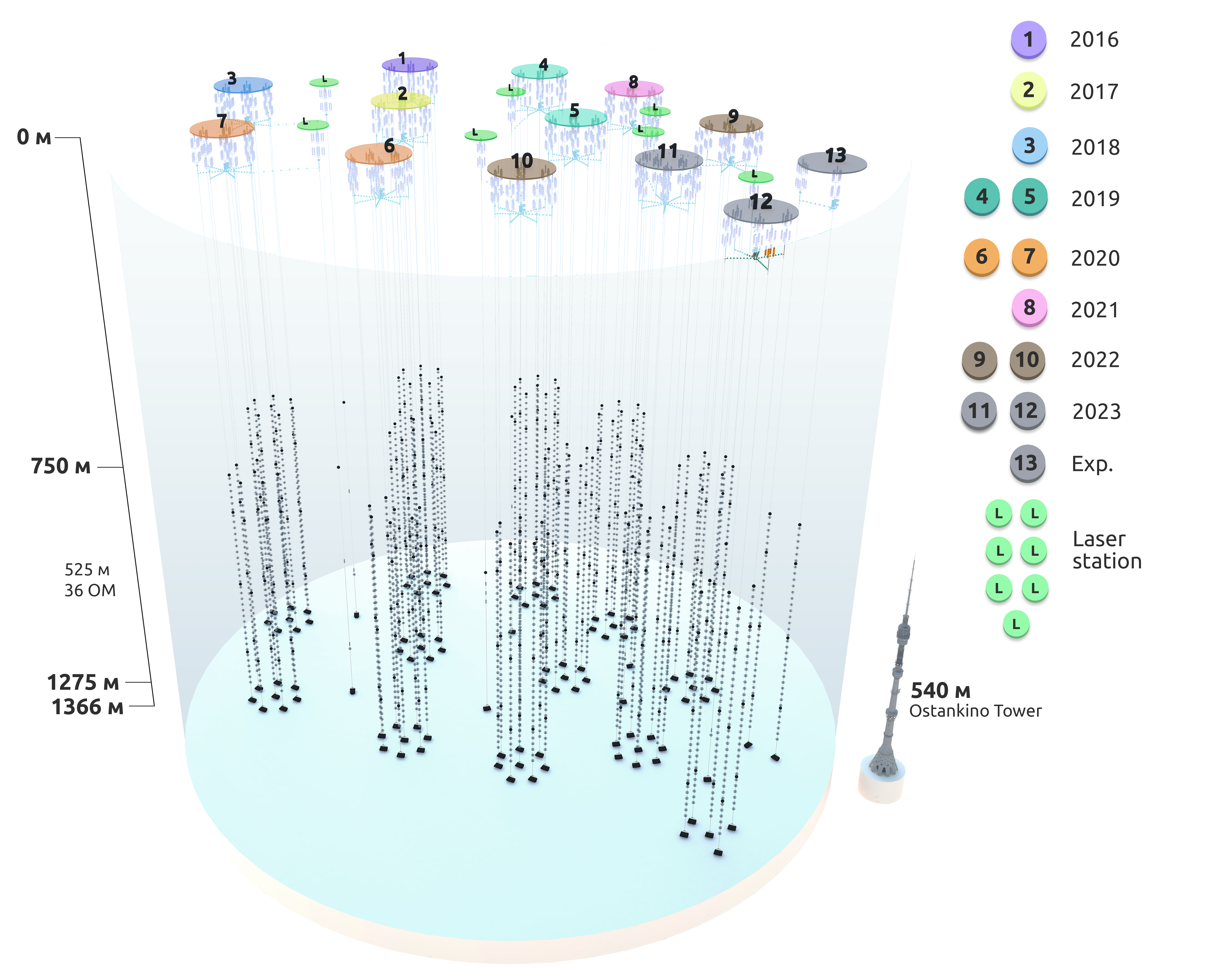}
			\label{obr:design}
		}   
	\end{multicols}

	\caption{a) Optical module of Baikal-GVD with 10" photomultiplier tube. b) Schematic view of Baikal-GVD neutrino telescope in 2023. In the legend, annual progression of detector installation is given.}
	\label{fig:Baikal_GVD}
\end{figure}

The main component of Baikal-GVD is a 10" photomultiplier tube enclosed in a pressure resistant glass sphere with a diameter of 42~cm -- optical module (OM) (see Fig.~\ref{obr:OM}). OMs are arranged on vertical structures called strings at depths of 750-1275~m. There are 36 OMs attached to one string with spacing of 15~m. The strings are organized in clusters. The cluster is a heptagonal, independently functioning unit consisting of 8 strings -- one central and seven peripheral strings. Distance between central string and peripheral strings is about 60~m. Central strings of neighboring clusters are separated by $\sim$300~m. After winter expedition in year 2023, Baikal-GVD consists of 3456 OMs installed on 96 strings (see Fig.~\ref{obr:design}).

According to the Cherenkov light topology, there are two main event signatures in Baikal-GVD -- cascades and tracks. It depends on the flavor of interacting neutrino which type of signature is formed. Muon events produce track-like signatures. Neutral current interactions of neutrinos of all flavors and charged current interactions of electron neutrinos create cascade-like Cherenkov light topologies.

In charged current interaction of tau neutrino, according to the $\tau$ decay mode, there are two types of Cherenkov light signatures that can be produced. In the case of $\tau$-lepton decay into muon, a cascade and a track signature is created. The branching ratio of this decay is approximately 17~\%~\cite{particle_data}. $\tau$-lepton decay into electron or hadrons (branching ratio $\sim$83\%~\cite{particle_data}) generates double cascade signature. 

The importance of tau neutrino detection stems from the fact that production rate of tau neutrinos in the atmosphere is negligible~\cite{Palladino:Importance_tau}. Therefore, if tau neutrino interaction is identified, there is a very high probability of astrophysical origin of this neutrino. In this paper, description of reconstruction technique for double cascade events, as well as evaluation of the performance of the algorithm is given.

\section{Double cascade reconstruction algorithm}
\begin{figure}[h!]
	\includegraphics[width=\linewidth]{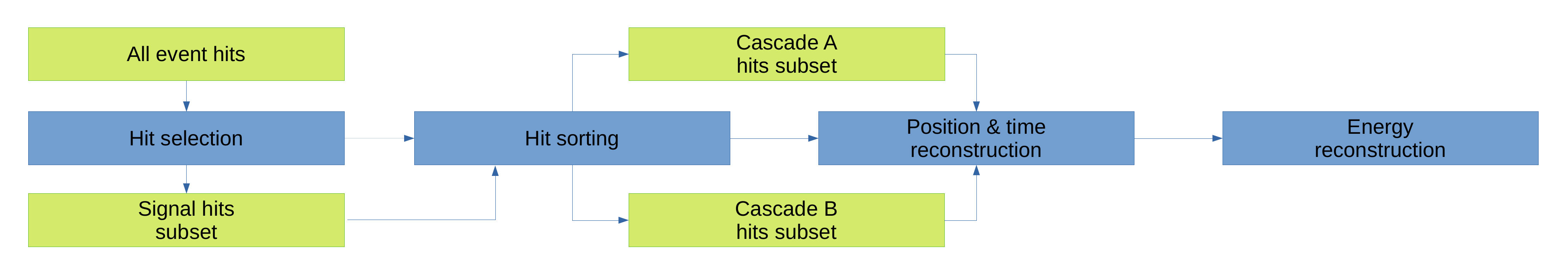} 
	\caption{Flowchart of the double cascade reconstruction algorithm.}
	\label{fig:DC_reco_flowchart}
\end{figure}

The main aim of the double cascade reconstruction algorithm is to determine parameters of double cascade events -- positions and times of cascade vertices, direction of double cascade event, and energies of both cascades. It is performed in four major steps -- hit selection, hit sorting, position and time reconstruction, and energy reconstruction (see Fig.~\ref{fig:DC_reco_flowchart}).

 This algorithm is an improved version of the reconstruction technique described in~\cite{ECRS2021:DC_Eckerova}.

\subsection{Hit selection}

The goal of the first step of the algorithm is to select signal hits from the cascades and suppress noise hits. At the beginning, a set of reference pulses is chosen. In current version of the algorithm 5 suitable pulses with the highest charges are tagged as reference pulses. Other hits are selected according to the condition:
\begin{equation}
	\mid T^{ref}-T^{meas}_{i} \mid <d_{i}/ v + \delta t,
	\label{eq:Causality}
\end{equation}
where $T^{ref}$ is detection time of the reference pulse, $T^{meas}_{i}$ is time of detection of studied pulse, $d_{i}$ is distance between OMs that detected these pulses, $v$ is the speed of light in water, and $\delta t$ is parameter that determines stringency of the criterion. For pulse to be selected, it needs to fulfill criterion given in Eq.~\ref{eq:Causality} with respect to one of the reference pulses. There is also additional criterion that one of the neighboring OMs (two above and two below) has to detect hit in a certain time window (usually at the level of 100~ns).

\subsection{Hit sorting}

The second step of the double cascade reconstruction algorithm is to categorize selected signal pulses to two groups -- one that corresponds to the cascade created in the $\nu_{\tau}$ interaction vertex and the second one that consists of pulses from $\tau$-lepton decay cascade.

Sorting of hits to the two subsets is performed using the criterion:
\begin{equation}
	\mid T^{meas}_{i}-T^{exp}_{i}(\vec{R},T) \mid \lesssim \delta T,
	\label{eq:tfilter}
\end{equation}
where $T^{meas}_{i}$ is time of detection of pulse, $T^{exp}_{i}$ is expected detection time of hit determined from the first estimated position $\vec{R}$ and time $T$ of the cascade vertex, and $\delta T$ defines strictness of the criterion. This procedure is repeated twice, each time for different cascade vertex ($\vec{R_{1}}$,$T_{1}$ and $\vec{R_{2}}$,$T_{2}$). The cascade is labeled as cascade A when its pulses are selected to the first subset, the other one is labeled as cascade B.

\begin{figure}[h!]
	\begin{multicols}{2}
		\subfloat[]{
			\includegraphics[width=\linewidth]{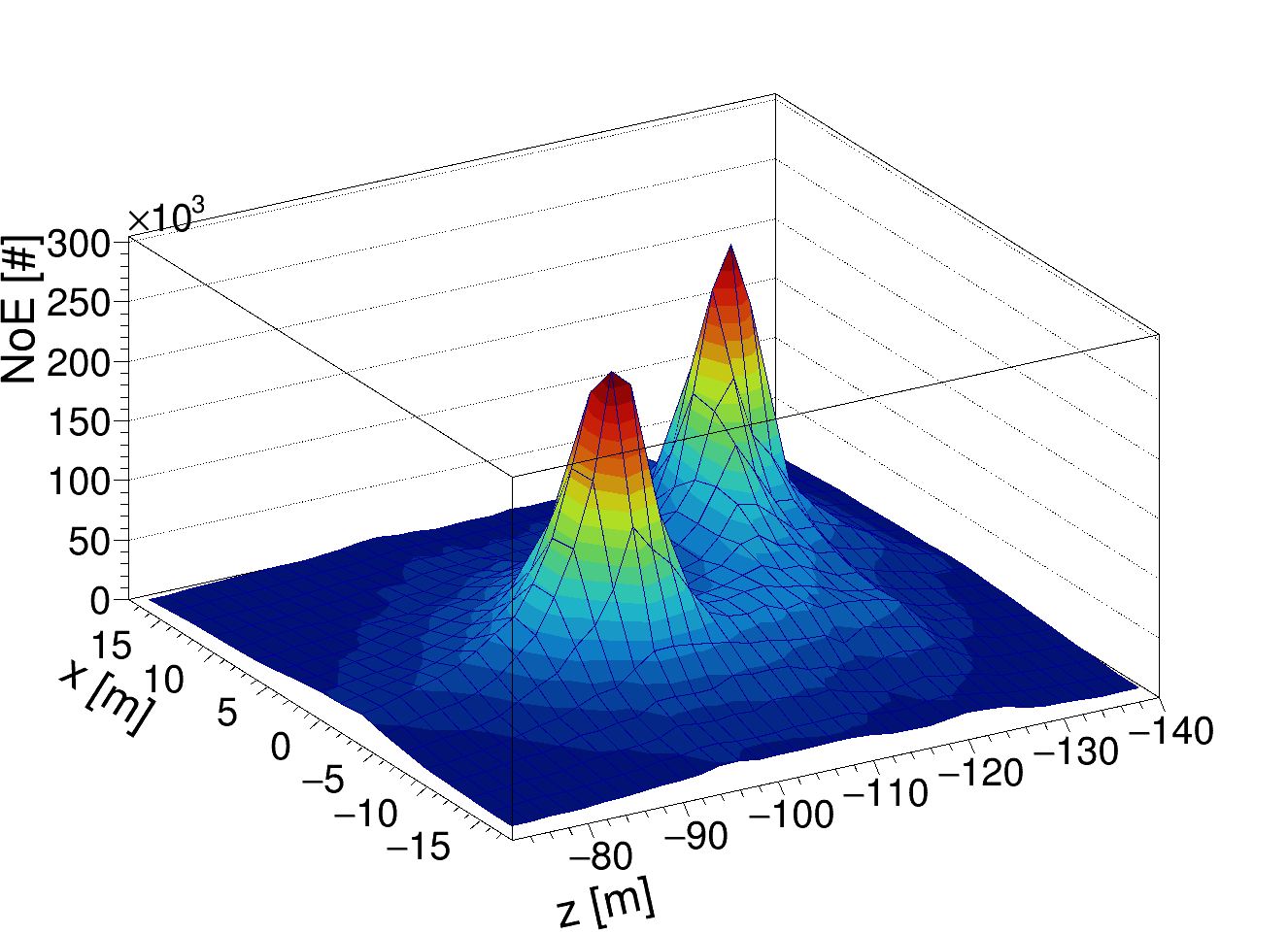} 
			\label{subfig:XZ_plane}
		}
		\subfloat[]{
			\includegraphics[width=\linewidth]{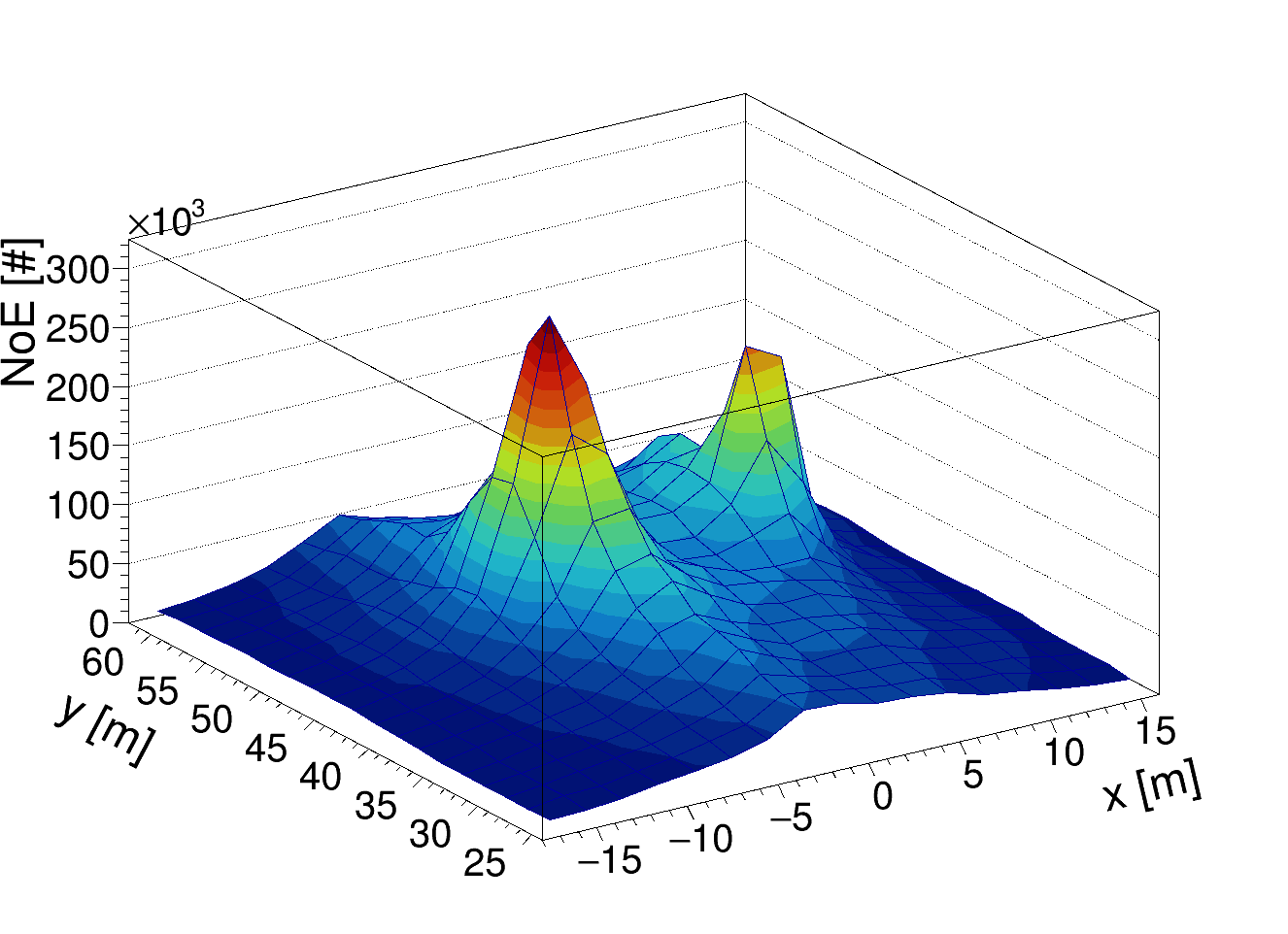}
			\label{subfig:XY_plane}
		}        
	\end{multicols}
	\caption{The display of the position estimations obtained by solving of Eq.~\ref{eq:estimate_pos_time}, using all sets of 5 pulses for $\nu_{\tau}$ event from MC simulations. From all position estimations, two peaks that correspond to two cascade vertices were formed. The position of $\nu_{\tau}$ cascade vertex is simulated at [9, 47, -126], and simulated position of $\tau$ decay cascade vertex is [-3, 45, -100]. a) XZ plane and b) XY plane of ($\vec{R}$, $T$)-space.}
	\label{fig:set_of_five_demonstration}
\end{figure}

The first estimated positions ($\vec{R_{1}}$, $\vec{R_{2}}$) and times ($T_{1}$, $T_{2}$) of the cascade vertices required in Eq.~\ref{eq:tfilter} are obtained by performing a scan of a space-time with several position and time estimations of cascade vertices (($\vec{R}$, $T$)-space). It is assumed that in the case of double cascade event, if several sets of pulses are used to estimate position and time of cascade vertex, two peaks in ($\vec{R}$, $T$)-space that corresponds to the two cascade vertices should be seen (see Fig.~\ref{fig:set_of_five_demonstration}). Afterwards, the scan procedure is able to identify these peaks and establish the first estimation of positions and times of the cascade vertices.

 Each position and time estimation used in the ($\vec{R}$, $T$)-space scan is determined from group of 5 pulses by solving a set of equations for distances $d_{i}$ between OMs that detected these pulses and cascade vertex:
\begin{equation}
	d_{i} = \sqrt{(x-x_{i})^{2}+(y-y_{i})^{2}+(z-z_{i})^{2}} = c/n(t-t_{i}),
	\label{eq:estimate_pos_time}
\end{equation}
where $\textbf{x} = (x,y,z,t)$ denotes space-time coordinates of the cascade vertex position, ${\textbf x_{i}} = (x_{i},y_{i},z_{i},t_{i})$ stands for space-time position of specific pulse registered on the OM, $c$ denotes speed of light in vacuum, and $n$ marks refractive index of water. The number of pulses (five) used for the estimation of the position and time of the cascade was selected because it is the lowest number of pulses needed to constrain $\textbf{x}$ unambiguously with this procedure~\cite{Hartmann:Thesis:2006}. Furthermore, the lowest possible number of pulses was chosen to increase probability of the pure set of pulses -- pulses that correspond to one cascade only. These sets enable to obtain the most precise estimation of position and time of cascade vertex.

\subsection{Position and time reconstruction}

The purpose of the third step of the double cascade reconstruction algorithm is to determine final values of the positions and times of the cascade vertices. The direction of the double cascade event is defined as a vector connecting the two cascade vertices. Aforementioned attributes of double cascade event are obtained by minimization of $\chi^2$ distribution:
\begin{equation}
	\chi^2 = \frac{1}{N_{hit1} + N_{hit2} - 7} \left(\sum\limits_{i = 0}^{N_{hit1}} \frac{ ( T_{1i}^{meas} - T_{1i}^{exp}(\vec{R_{1}},T_{1}))^2}{\sigma_t^2} + \sum\limits_{i = 0}^{N_{hit2}} \frac{ ( T_{2i}^{meas} - T_{2i}^{exp}(\vec{R_{2}},T_{2}))^2}{\sigma_t^2}\right).
	\label{eq:chi2_DC}
\end{equation}
In this formula $N_{hit1}$ and $N_{hit2}$ stand for numbers of hits in groups of pulses corresponding to the two cascades, $T_{1i}^{meas}$ and $T_{2i}^{meas}$ denote times of detection of pulses from the subsets on $i^{th}$ OM, $T_{1i}^{exp}$ and $T_{2i}^{exp}$ are expected detection times of pulses on $i^{th}$ OM determined with respect to the positions and times of cascade vertices ($\vec{R_{1}},T_{1}$ and $\vec{R_{2}},T_{2}$), and $\sigma_t$ means uncertainty in the time measurement~\cite{ICRC2021:cascades}. There are seven free parameters -- positions of both cascades ($\vec{R_{1}}$, $\vec{R_{2}}$) and time of the cascade A ($T_{1}$), time of the cascade B is calculated from aforementioned parameters.

\subsection{Energy reconstruction}
The last step of the algorithm is to determine energies of both cascades. There are two procedures for energy reconstruction -- energy prefit and minimization of log-likelihood formula. The energy prefit is a method in which energies of the cascades are obtained from particular pulse/s from known charge of the pulse, position and orientation of the OM that detected the pulse and already determined position and direction of the cascade.
The log-likelihood formula is given by equation:

\begin{equation}
	L =- \sum_{i=0}^{hit OM} \log (P_{i}(q_{i} \mid Q_{i})) - \sum_{i = 0}^{nonhitOM} \log (P_{i}(q_{i} = 0 \mid Q_{i})),
	\label{eq:likelihood}
\end{equation}
where $P_{i}(q_{i} \mid Q_{i})$ denotes the probability of detecting charge $q_{i}$ on $i^{th}$ OM while charge $Q_{i}$ is expected. The expected charge $Q_{i}$ is determined as a sum of the charges expected to be detected from both cascades:
\begin{equation}
	Q_{i} =
	Q(\vec{r_{1i}},\theta,\phi,E_{1})+Q(\vec{r_{2i}},\theta,\phi,E_{2}).
\end{equation}
This calculation takes into account positions and direction of the cascades, energies of the cascades, and position and orientation of the OM~\cite{ICRC2021:cascades}.

\section{Performance}

The performance of the double cascade reconstruction algorithm was evaluated using MC simulations of $\nu_{\tau}$ double cascade events. The expected astrophysical neutrino flux used in these simulations is given by the formula~\cite{PhysRevD:DiffuseFlux}:
\begin{equation}
	\phi(E) = 2.41\cdot10^{-5} \cdot \left(\frac{E}{\text{GeV}}\right)^{-2.58} [\text{GeV}^{-1}\text{s}^{-1}\text{sr}^{-1}\text{cm}^{-2}].
	\label{eq:astro_flux}
\end{equation}
Subsequent precision evaluations were performed using subset of all MC generated $\nu_{\tau}$ double cascade events -- double cascade like events. The main selection requirements are MC simulated distance between cascade vertices larger than 10~m and energy of $\nu_{\tau}$ higher than 100~TeV.

In Tab.~\ref{tab:Precision} the precision of reconstruction of cascade vertex positions is summarized. For particular parameters of double cascade events, mean and median values of distributions showing difference between reconstructed and simulated values of these parameters are presented.

\begin{table}[h!]
	\caption{The reconstruction precision of particular parameters of the double cascade event. The mean and median values of distributions of difference between simulated $x_{sim}$ and reconstructed $x_{reco}$ values of parameters of double cascade events are given. The results shown in this table are preliminary.}
	\label{tab:Precision}
	\centering
	\begin{tabular}{>{\centering\arraybackslash}p{4.6cm}||>{\centering\arraybackslash}p{1.2cm}|>{\centering\arraybackslash}p{1.2cm}}
		\hline
		Double cascade & \multicolumn{2}{c}{|$x_{sim} - x_{reco}$|} \\ 
		parameter  & mean & median    \\
		\hline \hline
		cascade A position [m]  & 2.76  & 2.21 \\ \hline
		cascade B position [m]  & 4.11  & 2.44 \\ \hline
		distance between vertices [m]    & 1.96  & 0.71 \\ \hline	
	\end{tabular}
\end{table}

\begin{figure}[H]
	\begin{center}
		\includegraphics[width=0.9\textwidth, trim={1cm 1cm 1cm 1cm},clip]{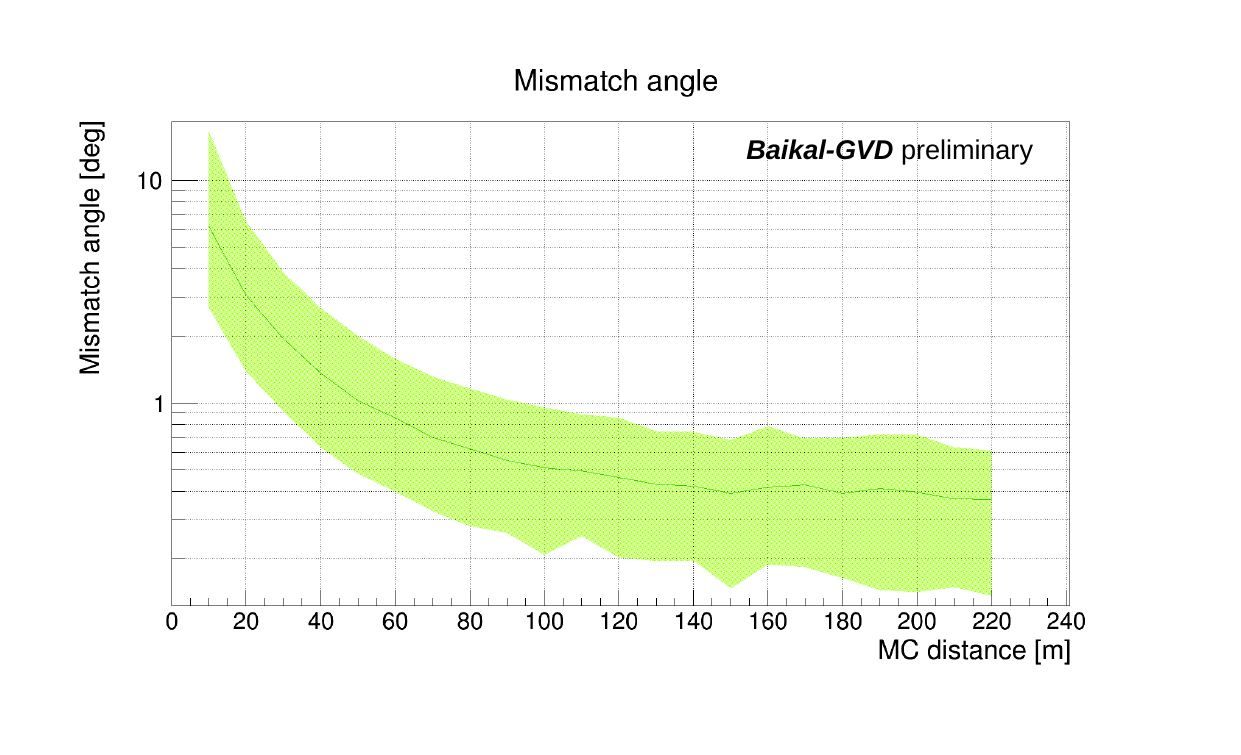}
		\caption{The median value (green line) and 68\% containment region (green belt) of distribution of angle between simulated and reconstructed direction of $\nu_{\tau}$ double cascade event as a function of true distance between cascade vertices.} 
		\label{fig:direction_reco}
	\end{center}
\end{figure}

In Fig.~\ref{fig:direction_reco} the angle between simulated and reconstructed direction of double cascade events is shown with respect to the simulated distance between cascade vertices. A green belt represents 68\% containment region of the distribution and the line inside this belt depicts the median of the distribution.

\begin{figure}[h!]
	\begin{center}
		\includegraphics[width=0.9\textwidth, trim={1cm 1cm 1cm 2cm},clip]{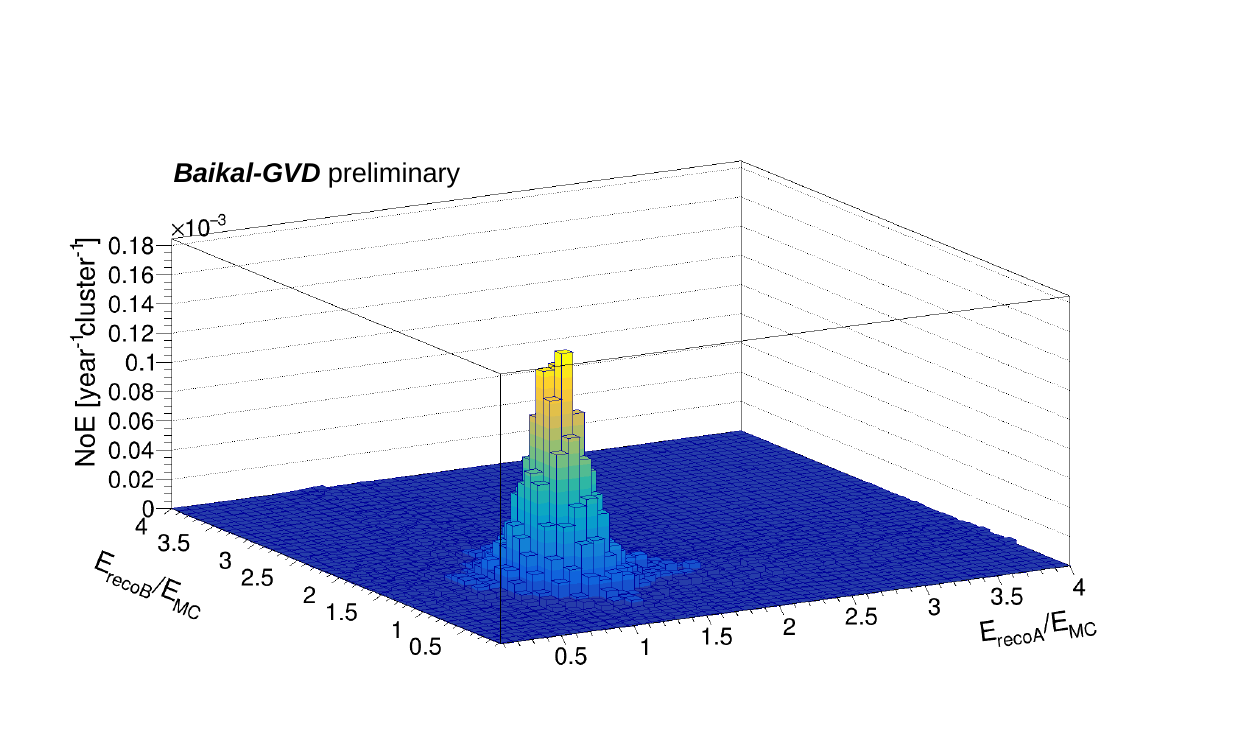}
		\caption{The distribution of fractional errors in energy reconstruction of double 
			cascade events ($\text{E}_{\nu_{\tau}}$ > 100 TeV; true distance between cascade vertices > 10 m; MC simulation).} 
		\label{fig:energy_reco}
	\end{center}
\end{figure}

The precision of energy reconstruction is summarized in Fig.~\ref{fig:energy_reco}. In this figure, the two-dimensional distribution of ratios of reconstructed to simulated energy of cascade A and B is displayed. This distribution was fitted with two-dimensional Gaussian function:
\begin{equation}
	f(x,y) = A \cdot e^{-\frac{(x-\mu_{A})^{2}}{2\sigma_{A}^{2}}-\frac{(y-\mu_{B})^{2}}{2\sigma_{B}^{2}}}.
\end{equation}
 The parameters obtained from the fit are $\mu_{A}$ = 1.02, $\mu_{B}$ = 1.04, $\sigma_{A}$ = 0.18, and $\sigma_{B}$ = 0.24.

\section{Conclusion}
In this paper, further steps in the development of the double cascade reconstruction technique in Baikal-GVD were presented. The description of the current version of this algorithm was given. The precision of the reconstruction of the double cascade event parameters, particularly positions of the cascade vertices, distance between cascade vertices, direction of the double cascade event, and energies of the cascades was presented. An extension of the double cascade reconstruction technique for events with distance between cascade vertices shorter than 10~m is in development.

\end{document}